\documentclass[article,pra,aps,twocolumn]{revtex4}
\usepackage{graphicx}
\begin{document}
\title{Ranking Scientific Publications Using a Simple Model of Network Traffic}
\author{Dylan Walker$^{1,2}$, Huafeng Xie$^{2,3}$, Koon-Kiu Yan$^{1,2}$, Sergei Maslov$^2$}
\affiliation{
$^1$Department of Physics and Astronomy, Stony Brook University, \\
Stony Brook, New York, 11794, USA\\
$^2$Department of Condensed Matter Physics and Materials Science, Brookhaven National Laboratory,
Upton, New York 11973,  USA\\
$^3$New Media Lab, The Graduate Center, CUNY, \\
New York, NY 10016, USA}
\date{\today}
\begin{abstract}
To account for strong aging characteristics of citation networks, we modify Google's PageRank algorithm by initially distributing random surfers exponentially with age, in favor of more recent publications. The output of this algorithm, which we call CiteRank, is interpreted as approximate traffic to individual publications in a simple model of how researchers find new information. We develop an analytical understanding of traffic flow in terms of an RPA-like model and optimize parameters of our algorithm to achieve the best performance. The results are compared for two rather different citation networks: all American Physical Society publications and the set of high-energy physics theory (hep-th) preprints. Despite major differences between these two networks, we find that their optimal parameters for the CiteRank algorithm are remarkably similar.
\end{abstract}
\maketitle

Due to their rapid growth and large size, many information networks have become untenable to navigate without some sort of ranking scheme. This is particularly evident in the example of the World Wide Web, a network of pages connected by hyperlinks. A successful solution to the problem of ranking the Web is Google's PageRank algorithm \cite{PageBrin}. Another class of information networks that could benefit from such a ranking method are citation networks. These networks are comprised of scientific publications connected by citation links.

Current methods of ranking publications based on the total number of citations received are rather crude. They are too ``democratic'' in treating all citations as equal and ignoring differences in importance of citing papers. One of the advantages of Google's PageRank algorithm is that it implicitly accounts for the importance of the citing article in a self-consistent fashion. Authors of \cite{google_citation_bolen} proposed using the PageRank algorithm to improve the formula used to calculate the impact factor of scientific journals. In \cite{Pu_Redner_2006} some of us directly applied this algorithm to individual papers published in all American Physical Society journals.
This allowed us to discover a set of highly influential papers (``scientific gems") that would be undervalued based on just their number of citations. However, there exist significant differences between the World Wide Web and citation networks that suggest a modification of the original PageRank algorithm. The most important difference is that, unlike hyperlinks, citations cannot be updated after publication.   This makes aging effects \cite{Price,aging1} in citation networks much more pronounced than in the WWW. The other consequence is the inherent time-arrow present in the topology of citation networks, due to the constraint that a paper may only cite earlier works. This significantly alters the spectral properties of the adjacency matrix which lie at the heart of the PageRank algorithm. In particular, the absence of directed loops means that the adjacency matrix can have only zero eigenvalues.

The success of the PageRank algorithm can be attributed, in part, to its ability to capture the behavior of people randomly browsing the network of web pages. Indeed, the PageRank of a given web page can be interpreted as the predicted traffic (quantified e.g., by the rate of downloads) for that page if every WWW user follows a random path of (on average) $1/\alpha$ hyperlinks starting from a randomly selected webpage. The assumption that a typical web-surfer starts at a randomly selected webpage might be not completely unreasonable for the WWW, but it needs to be modified for citation networks. As all of us know, researchers typically start ``surfing" scientific publications from a rather {\it recent} publication that caught their attention on a daily update of a preprint archive or a recent volume of a journal. Thus a more realistic model for the traffic along the citation network should take into account that researchers preferentially start their quests from recent papers and progressively get to older and older papers with every step.

In this work we introduce the CiteRank algorithm, an adaptation of the PageRank algorithm to citation networks. Our algorithm simulates the dynamics of a large number of researchers looking for new information. Every researcher, independent of one another, is assumed to start his/her search from a \textit{recent} paper or review and to subsequently follow a chain of citations until satisfied. Explicitly, we define the following two-parameter CiteRank model of such a process, allowing one to estimate the traffic $T_i(\tau_{dir} ,\alpha)$ to a given paper $i$. A recent paper is selected randomly from the whole population with a probability that is exponentially discounted according to the age of the paper, with a characteristic decay time of $\tau_{dir}$.
At every step of the path, with probability $\alpha$ the researcher is satisfied and halts his/her line of inquiry.  With probability $(1-\alpha)$ a random citation to an adjacent paper is followed. The predicted traffic, $T_i(\tau_{dir},\alpha)$, to a paper is proportional to the rate at which it is visited if a large number of researchers independently follow such a simple-minded process.

While we interpret the output of the CiteRank algorithm as the traffic, its utility ultimately lies in the ability to successfully rank publications. High CiteRank traffic to a publication denotes its high relevance in the context of currently popular research directions, while the PageRank number is more of a ``lifetime achievement award'' \cite{Pu_Redner_2006}. It is fruitful to compare the CiteRank traffic to a paper, $T_i$, with the more traditional method of ranking publications, the number of citations received.  Indeed, the two are highly correlated; a result easily understood on the basis that the larger the number of citations a paper has, the more likely it will be visited by a researcher via one of the incoming links.

However, the more refined CiteRank algorithm surpasses the conventional ranking, by number of citations, in its characterization of relevancy on two accounts.  Like the original PageRank algorithm \cite{PageBrin}\cite{google_citation_bolen}, in CiteRank, the popularity of papers is calculated in a self-consistent fashion: The effect of a citation from a more popular paper is greater that that of a less popular one.  A citation from a paper that is ``highly visible" will contribute more to the visibility of the cited paper. Furthermore, the age of a citing paper is intrinsically accounted for. The effect of a recent citation to a paper is greater than that of an older citation to the same paper.  New citations indicate the relevancy of a paper in the context of current lines of research.

An algorithmic description of the aforementioned model can be understood as follows. The transfer matrix associated with the citation network is $W_{ij} = 1/k^{out}_j$ if $j$ cites $i$ and 0 otherwise, where $k^{out}_j$ is the out-degree of the jth paper. Let $\rho_i$, the probability of initially selecting the $i^{th}$ paper in a citation network, be given by $\rho_i= e^{-age_i/\tau_{dir}}$.  The probability that the researcher will encounter a paper by initial selection alone is given by $\vec{\rho}$. Similarly, the probability of encountering the paper after following one link is $(1-\alpha)W\cdot\vec{\rho}$. The CiteRank traffic of the paper is then defined as the probability of encountering it via paths of any length:

\begin{equation}
\vec{T} = I\cdot\vec{\rho}  + (1-\alpha)W\cdot\vec{\rho} +
(1-\alpha)^2W^2\cdot\vec{\rho} + \cdots
%
\label{iteration}
\end{equation}
Practically, we calculate the CiteRank traffic on all papers in our dataset by taking successive terms in the above expansion to sufficient convergence ($<10^{-10}$ of the average value).

In order to assess the viability of this ranking scheme and to select optimal parameters $(\tau_{dir},\alpha)$, we need a quantitative measure of its performance on real citation networks. Two real citation networks are evaluated. \textbf{Hep-th}: An archive snapshot of the ``high energy physics theory'' archive from April 2003 (preprints ranging from 1992 to 2003). This dataset, containing around 28,000 papers and 350,000 citation links, was downloaded from \cite{hep_th}. \textbf{Physrev}: Citation data between journals published by the American Physical Society \cite{APS}. This dataset contains around 380,000 papers and 3,100,000 citation links ranging from 1893 to 2003.

Of course, evaluating the performance of any ranking scheme is a delicate, but often necessary, matter.  One way to select the best performing $\alpha$ and $\tau_{dir}$ is to optimize the correlation between the predicted traffic, $T_i(\tau_{dir},\alpha)$ and the actual traffic (e.g., downloads). Unfortunately, the actual traffic data for scientific publications are not readily available for these networks. However, it is reasonable to assume that traffic to a paper is positively correlated with the number of new citations it accrues over a recent time interval, $\Delta k_{in}$. For lack of better intuition we first assume a linear relationship between actual traffic and number of recent citations accrued. This corresponds to a simple-minded scenario in which every researcher downloading a paper will, with a small probability, add it to the citation list of the manuscript he/she is writing \cite{nomodel}. In order to compare CiteRank with actual citation accrual, we constructed an historical snapshot of the networks. In both cases, the most recent 10 percent of papers are pruned from the network. The CiteRank traffic, $T_i$, of the remaining 90 percent of the papers is then evaluated and correlated with their actual accrual of new citations, $\Delta k_{in}$, originating at the most recent 10 percent of papers.
\begin{figure}[htb]
\includegraphics*[width=\linewidth]{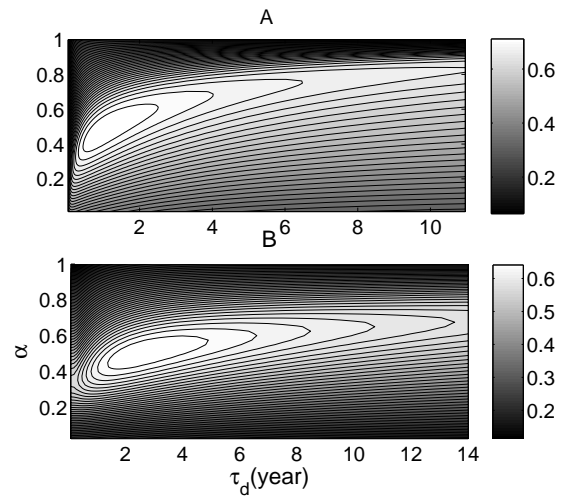}
\caption{
The Pearson (linear) correlation coefficient between the number of recent citations accrued ($\Delta k_{in}$) and CiteRank traffic ($T_i$) is calculated over the parameter space of the CiteRank model for the hep-th (A) and physrev (B) network. Both networks exhibit peaks in correlation coefficient in the $\alpha$-$\tau_{dir}$ plane.
The highest correlation is achieved for $\alpha=0.48$, $\tau_{dir}=1$ year in the hep-th network and 
$\alpha=0.50$, $\tau_{dir}=2.6$ years, in the physrev network.}
\label{fig1}
\end{figure}
It is important to note the qualitative and quantitative differences between the two citation networks considered. The Physical Review citation network (physrev) is comprised of a large number ($\sim$ 400,000) of peer-reviewed publications acquired over a period close to a hundred years.  The high-energy physics archive citation network (hep-th) is comprised completely of a much smaller number ($\sim$ 28000) of electronically submitted publication preprints, with no associated form of peer review.
Despite these significant differences in the nature of the networks considered, the general features of their correlation contours are outstandingly similar. In both cases, a single sharp peak in correlation is evident for particular values of the parameters. The value of the optimal parameters for both networks are:
\renewcommand{\labelitemi}{}
\begin{itemize}
    \item hep-th: $\alpha=0.48$, $\tau_{dir}=1$ year 
    \item physrev: $\alpha=0.50$, $\tau_{dir}=2.6$ years
\end{itemize}
Remarkably, the value of $\alpha$ is nearly the same for two rather different networks considered here and is in agreement with that proposed in \cite{Pu_Redner_2006} on purely empirical grounds.  The difference in optimal parameter $\tau_{dir}$ for these networks is in agreement with the common-sense expectation of faster response time (and hence faster aging of citations) in preprint archives compared to peer-reviewed publications.
Another feature of Fig. \ref{fig1} is that, in both networks, large values of the correlation coefficient are concentrated along a diagonally-positioned ridge.
In other words, the best choice of $\alpha$ for a given $\tau_{dir}$ seems to rise linearly with $\tau_{dir}$, a behavior that will be revisited later in this text.  The resultant CiteRank traffic and corresponding ranking for the two citation networks can be accessed here \cite{onlineranking}.

\begin{figure}[htb]
\includegraphics*[width=\linewidth]{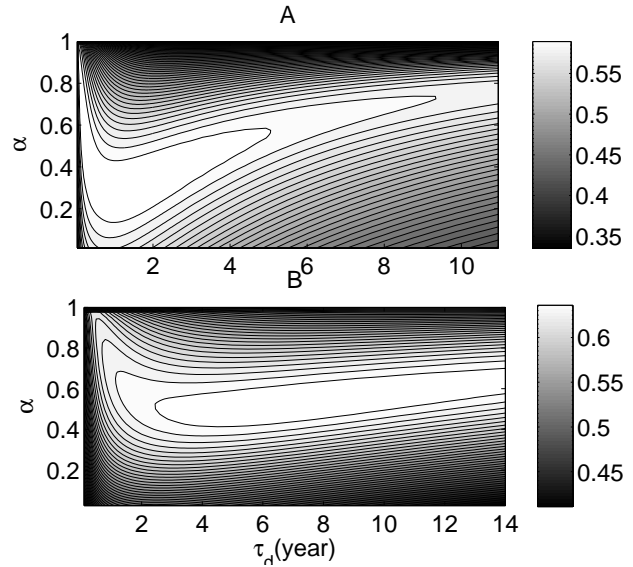}
\caption{
The Spearman rank correlation coefficient between recent citations accrued ($\Delta k_{in}$)
and CiteRank traffic ($T_i$) for the
hep-th (A) and physrev (B) network.  Both networks exhibit similar behavior.
There are more extended regions of good correlation relative to the linear correlation
contours of fig. 1.  This broadening is expected as a consequence of the more relaxed correlation measure.
The highest rank correlation occurs for
$\alpha=0.31$, $\tau_{dir}=1.6$ years, in the hep-th network and
$\alpha=0.55$, $\tau_{dir}=8$ years, in the physrev network.
}\label{fig2}
\end{figure}

While the correlation contour plots shown in Fig. \ref{fig1} are a promising indication that the CiteRank model of traffic provides a good zero-order approximation to the actual traffic along a citation network, they are to some extent predicated on the assumption of a linear relationship between actual traffic and $\Delta k_{in}$. One might readily ask how this model fares in the absence of such an assumption. While the assumption of a {\it linear} relationship may be unreasonable, a positive, monotonic relationship between these quantities is certainly expected.  There is a statistical correlation method precisely adapted for such a situation, namely, the Spearman rank correlation.  Under this relaxed correlation measure, only the rank of $T_i$ are correlated with the rank of $\Delta k_{in}$. Numerical changes in $T_i$ that do not lead to reordering have no effect on the value of the rank correlation coefficient. Another rationale for using rank correlations is that our ultimate goal is ranking publications, not modeling the traffic. Thus, we are currently not interested in individual $T_i$'s,  but only in their relative values. Spearman correlation contour plots are constructed for both networks and shown in Fig. \ref{fig2}. The optimal values for both networks are:
\renewcommand{\labelitemi}{}
\begin{itemize}
    \item hep-th: $\alpha=0.31$, $\tau_{dir}=1.6$ year 
    \item physrev: $\alpha=0.55$, $\tau_{dir}=8$ years
\end{itemize}
These results roughly confirm the prediction of $\alpha\sim0.5$ from Fig. \ref{fig1}, however there is a more appreciable discrepancy in $\tau_{dir}$ between linear and rank correlation for both networks.

In both panels of Fig. \ref{fig1}, over a broad range of parameters, the optimal value of $\alpha(\tau_{dir})$ for a given value of $\tau_{dir}$ is positively correlated with $\tau_{dir}$.  This is an indication that these two parameters are entangled. In fact, this is to be expected as it is some admixture of the two parameters which leads to the exposure of a given paper to the researcher. An intuitive picture of this entanglement can be understood in terms of the penetration depth, which is a measure of how far back in time a random surfer following rules of the CiteRank algorithm is likely to get. The penetration depth is affected by both $\tau_{dir}$ - the average age of the initial paper at which he/she started following the chain of citations, and $1/\alpha$ - the mean number of steps on this chain of citations. For small $\tau_{dir}$ and large $\alpha$, the penetration depth is small, implying that only very recent papers receive traffic. On the other hand, for large $\tau_{dir}$ and small $\alpha$, the penetration depth is very large, indicating that most of the traffic is directed towards older papers.

\begin{figure}[htb]
\includegraphics*[width=\linewidth]{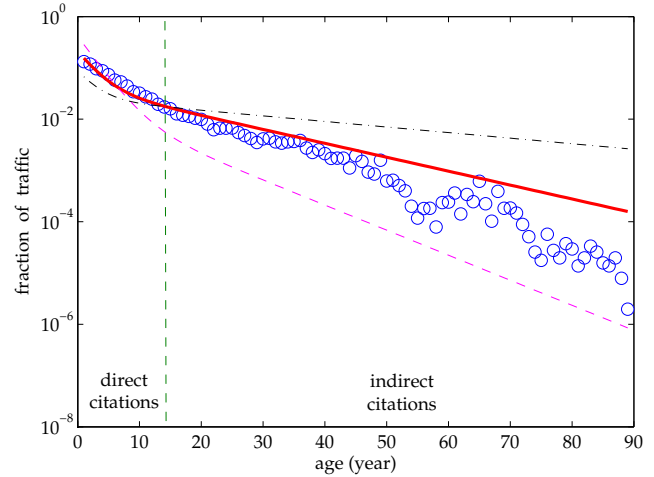}
\caption{The age distribution of  newly accrued citations $\Delta k_{in}$ (blue) for the physrev network. Theoretical predictions [\ref{cross_over}] for the CiteRank traffic are calculated for the optimal $\tau_{dir}=2.6$ and three values of $\alpha=0.2$ (dot-dashed line), $0.5$ (thick solid line), and $0.9$ (dashed line). In agreement with Fig.1, the optimal value, $\alpha=0.5$, provides the best agreement with $\Delta k_{in}$. All curves are normalized so that the sum of all data points is equal to 1.}\label{fig3}
\end{figure}

To better understand how $\alpha$ and $\tau_{dir}$ influence the age distribution of CiteRank traffic, we performed the following quantitative analysis. Let $T_{tot}(t)$ denote the total CiteRank model traffic to papers written exactly $t$ years ago. As described by Eq. \ref{iteration}, two distinct processes contribute to $T_{tot}(t)$. The first is the ``direct'' traffic $T_{dir}(t)$ due to the initial selection of papers in this age group, which is proportional to $\exp(-t/\tau_{dir})$ \cite{caveat}. The second is the ``indirect'' traffic $T_{ind}(t)$ arriving via one of the incoming citation links.  The latter is given by $T_{ind}(t)=(1-\alpha)\int_t^\infty T_{tot}(t')P_c(t',t)dt'$, where $P_c(t',t)$ is the fraction of citations originating from papers of age $t'$ that cite papers of age $t$. It should be noted that $P_c(t',t)$ is an \textit{empirical} distribution and, as such, is a \textit{measured} property of the citation network under consideration.  According to \cite{aging1} and our own findings, $P_c(t',t)$ is reasonably well approximated by the exponential form $\frac{1}{\tau_c}\exp(-(t'-t)/\tau_c)$. Taking the Fourier transform of the equation $T_{tot}(t)=T_{dir}(t)+T_{ind}(t)$, we have
\begin{equation}
T_{tot}(\omega)=T_{dir}(\omega)+(1-\alpha)T_{tot}(\omega)P_c(\omega).
\label{t_omega}
\end{equation}
This equation is similar, in spirit, to the well-known random phase approximation \cite{RPA}.  Solving Eq. \ref{t_omega} for $T_{tot}(\omega)$ and taking the inverse Fourier
transform, yields
\begin{equation}
T_{tot}(t)\sim (\tau_c-\tau_{dir})\exp(-t/\tau_{dir})+(1-\alpha)\tau_{dir}\exp(-\alpha t/\tau_c).
\label{cross_over}
\end{equation}
Thus, the traffic arriving at the subset of papers of age $t$ is given by the superposition of two exponential functions. We are now in a position to better understand what determines the optimal values of $\alpha$ and $\tau_{dir}$. Open circles in Fig. \ref{fig3} show the age distribution of the number of recently acquired citations, $\Delta k_{in}$, for papers in the physrev dataset. The approximate CiteRank traffic, given by Eq. \ref{cross_over}, is also displayed. It is calculated using the empirically determined value $\tau_c=8$ years, optimal $\tau_{dir}=2.6$ years and three values of $\alpha=0.2,0.5$ and $0.9$. As one would expect, the profile of $\langle \Delta k_{in} \rangle$ vs $t$ best agrees with the CiteRank plot for the optimal value $\alpha=0.5$ \cite{war}. Fig. \ref{fig3} also provides some clues to the positive correlation between near-optimal choices of $\alpha$ and $\tau_{dir}$, visible as diagonal ``ridges" in Fig. \ref{fig1}A and B. Indeed, if the value of $\alpha$ is chosen to be large, the contribution from the second term is diminished; the use of a larger value of $\tau_{dir}$ could partially compensate for the loss of  CiteRank traffic to older papers, and would thus be in reasonably good agreement with the $\Delta k_{in}$ data.

Another encouraging observation is that, like Eq. \ref{cross_over}, the age distribution of recently acquired citations shown in Fig. \ref{fig3} has two regimes characterized by two different decay constants of about $5$ and $16$ years, with a crossover point around $t=15$ years. Our interpretation of this fact is that papers are found and  cited via two distinct mechanisms: researchers can either find a paper directly or by following citation links from earlier papers. For each of these mechanisms, the probability that a given paper is found decays with its age but the characteristic decay time for the direct discovery is shorter. While very recent papers, especially the ones altogether lacking citations, are for the most part discovered directly, older papers are mostly discovered by following citation links.

The optimal values of $\alpha$ in the two very different citation networks considered are remarkably close to each other. In both cases it appears that, on average, the length of chains of citations followed by a typical researcher is close to $1/\alpha \simeq 2$. Since this chain includes the original starting point, the length of around 2 means that the average cited paper is just one link away from the starting point. This raises the disconcerting possibility that many citations may be copied directly from the initially discovered reference.  Such citation copying was recently proven to be a very common scenario \cite{sirkin}.

Work at Brookhaven National Laboratory was carried out under Contract No. DE-AC02-98CH10886, Division of Material Science, U.S. Department of Energy. We are extremely grateful to Mark Doyle, Marty Blume, and Paul Dlug from the Physical Review Editorial Office for providing us with the APS citation data used in this work.  We thank S. Redner and P. Chen for helpful discussions.


\begin{thebibliography}{5}
\bibitem{PageBrin} S. Brin and L. Page, 
Computer Networks and ISDN Systems, {\bf 30}, 107 (1998).
%
\bibitem{google_citation_bolen}
J. Bollen, M. A. Rodriguez, and H. Van de Sompel
cs.DL/0601030
%
\bibitem{Pu_Redner_2006} P. Chen, H. Xie, S. Maslov, S. Redner, physics/0604130
%
\bibitem{Price} D. J. De Solla Price, Science {\bf 149}, 510
(1965). 
%
\bibitem{aging1} S. Redner, Physics Today {\bf 58}, 49 (2005)
%
\bibitem{hep_th} This hep-th dataset was used in the KDD Cup 2003
\begin{verbatim}http://www.cs.cornell.edu/projects/kddcup/.\end{verbatim}
%
\bibitem{APS} The APS journals include Phys. Rev. Series I (1893–-1912), Series II (1913–- 1969), and Series III (1970–-present). This latter series includes the five topical sections: Phys. Rev. A, B, C, D, and E. Also included are Phys. Rev. Lett., Rev. Mod. Phys., and Phys. Rev. Special Topics, Accelerators and Beams.
%
\bibitem{nomodel} It should be noted that we make no attempt to model network growth in this paper.
%
\bibitem{onlineranking} \begin{verbatim}http://cmth.bnl.gov/~maslov/citerank/\end{verbatim}
%
\bibitem{RPA} J. Jensen, A. Mackintosh, Rare Earth Magnetism: Structures and Excitations, 155, Clarendon Press, Oxford, 1991.
%
\bibitem{caveat} Precisely speaking $T_{dir}(t)$ in the CiteRank model
is given by $N_p(t)$ -- the number of papers of age $t$ -- multiplied by the
exponential probability of selection $\exp(-t/\tau_{dir})$.
Since $N_p(t)$ itself often has approximately
exponential form with time constant $\tau_p$,
$\tau_{dir}$ used in the following equations should be
``renormalized'' to $\tilde{\tau_{dir}}=\tau_{dir} \cdot
\tau_p/(\tau_p+\tau_{dir})$. However, $\tau_p$ is
usually rather large ($\sim 28$ years in the PhysRev network). Thus
except for very large $\tau_{dir}$'s this
renormalization can be safely ignored.
%
\bibitem{sirkin} M.V. Simkin, V.P. Roychowdhury, Complex
Syst. {\bf 14}, 269 (2003).


\bibitem{war} The apparent disagreement in the tail involves profound dips due to the World War II and I \cite{aging1}, which of course cannot be explained by any theoretical model.

%
\end{thebibliography}
\end{document}